\documentstyle[prl,preprint,aps,epsf]{revtex}

\begin{document}
\bibliographystyle{plain}
\title{Condition for the Superradiance Modes in Higher-Dimensional Rotating
Black Holes with Multiple Angular Momentum Parameters}
\author{Eylee Jung\footnote{Email:eylee@kyungnam.ac.kr}, 
SungHoon Kim\footnote{Email:shoon@kyungnam.ac.kr} and
D. K. Park\footnote{Email:dkpark@hep.kyungnam.ac.kr 
}}
\address{Department of Physics, Kyungnam University,
Masan, 631-701, Korea.}
\date{\today}
\maketitle

\begin{abstract}
The condition for the existence of the superradiance modes is derived for the
incident scalar, electromagnetic and gravitational waves when the spacetime 
background is a higher-dimensional rotating black hole with multiple 
angular momentum parameters. The final expression of the condition is 
$0 < \omega < \sum_i m_i \Omega_i$, where $\Omega_i$ is an angular 
frequency of the black hole and, $\omega$ and $m_i$ are the energy of the 
incident wave and the $i$-th azimuthal quantum number. The physical implication
of this condition in the context of the brane-world scenarios is 
discussed.
\end{abstract}
\newpage
Recently, much attention is paid to the various properties of the absorption 
and emission problems in the higher-dimensional black holes. This is mainly 
due to the fact that the brane-world scenarios\cite{ark98-1,anto98,rs99-1}
opens the possibility to make tiny black holes in the future 
colliders\cite{gidd02-1,dimo01-1,eard02-1,stoj04} by high-energy scattering.
For the non-rotating black holes the complete absorption and emission spectra
are calculated numerically in Ref.\cite{jung05-2}, where the effect of the 
number of extra dimensions $n$ and the inner horizon radius $r_-$ on the 
spectra is carefully examined. It has been found in Ref.\cite{jung05-2} that 
the presence of $r_-$ generally enhances the absorptivity and suppresses 
the emission rate while the presence of $n$ reduces the absorptivity and 
increases the emission rate regardless of the brane-localized and bulk fields. 
In fact, this fact can be deduced by considering the Hawking temperature or 
by computing the effective potential generated by the horizon structure.

Also, the ratio of the low-energy absorption cross section for the Dirac 
fermion to that for the scalar field is derived analytically\cite{jung05-1}
in the charged black hole background. For the case of the bulk fields this
ratio factor becomes 
\begin{equation}
\label{bulk-ratio}
\gamma^{BL} \equiv \frac{\sigma_F^{BL}}{\sigma_S^{BL}} = 
2^{- (n+3)/(n+1)} \left[ 1 - \left( \frac{r_-}{r_+}\right)
^{n+1} \right]^{(n+2)/(n+1)}
\end{equation}
and for the case of the brane-localized fields this factor becomes
\begin{equation}
\label{brane-scalar}
\gamma^{BR} \equiv \frac{\sigma_F^{BR}}{\sigma_S^{BR}} =
2^{(n-3)/(n+1)} \left[ 1 - \left( \frac{r_-}{r_+}\right)
^{n+1} \right]^{2/(n+1)}.
\end{equation} 
In the Schwarzschild limit $(r_- \sim 0)$ $\gamma^{BL} \sim 2^{- (n+3)/(n+1)}$
and $\gamma^{BR} \sim 2^{(n-3)/(n+1)}$ which reduces to $1/8$ when $n=0$, which
was derived by Unruh long ago\cite{unruh76}. It is interesting to note that
$\sigma_F^{BL} = \sigma_S^{BL} / 2$ and $\sigma_F^{BR} = 2 \sigma_S^{BR}$ when
$n = \infty$. In the extremal limit $(r_- \sim r_+)$ the low-energy absorption 
cross sections for the brane-localized and bulk fermions goes to zero. The 
explicit $n$-dependence of these ratio factors may play important role in the 
experimental proof on the existence of the extra dimensions.

Recently, there is a controversy in the question of whether the 
higher-dimensional black holes radiate mainly on the brane or in the bulk.
Ref.\cite{argy98,banks99} argued that the Hawking radiation into the bulk is 
dominant compared to the brane-emission. The main reason for this is because
of the fact that for the tiny black holes the Hawking temperature is much 
larger than the mass of the light Kaluza-Klein modes. However, Ref.\cite{emp00}
argued that the Hawking radiation on the brane is dominant because the 
radiation into the bulk by the light Kaluza-Klein modes is strongly suppressed
by the geometrical factor. This argument is supported numerically by 
Ref.\cite{jung05-2} when $n$ is not too large in the charged black holes.

However, the situation can be completely different when the black hole 
background has an angular momentum. For the rotating black holes the incident 
waves can be scattered backward with extraction of the black hole's rotating
energy, which is called superradiance. The effect of the superradiance in the 
$4$-dimensional Kerr black hole was discussed in 
Ref.\cite{zeldo71-1,press72,star73-1,star74-1}. In this context 
Ref.\cite{frol02-1,frol02-2} argued that the conventional claim that 
{\it the black holes radiate mainly on the brane} can be changed if the effect
of the superradiance is involved. In fact, the existence the superradiance was
proved analytically\cite{frol03-1} and numerically\cite{ida05,harris05-1}. Also
the general condition for the existence of the superradiance modes for the 
scalar, electromagnenic and gravitational waves was derived in 
Ref.\cite{jung05-3} using the Bekenstein argument\cite{beken73}
when the black hole has a single angular momentum parameter. 

Here, we would like to extende Ref.\cite{jung05-3} to the 
rotating black holes
which have multiple angular momentum parameters. This is important because the 
tiny rotating black holes that will be produced in the future colliders due
to the nonzero impact parameter can have multiple components of the 
angular momentum since the brane thickness is of order of $1$/TeV. 

We start with a spacetime of the $(N+1)$-dimensional rotating black hole
derived by Myers and Perry in Ref.\cite{myers86}:
\begin{equation}
\label{spacetime1}
ds^2 = -dt^2 + \sum_{i=1}^{N/2} (r^2 + a_i^2) (d \mu_i^2 + \mu_i^2 d \phi_i^2)
+ \frac{\mu r^2}{\Pi {\cal F}}
\left( dt + \sum_{i=1}^{N/2} a_i \mu_i^2 d \phi_i \right)^2 +
\frac{\Pi {\cal F}}{\Pi - \mu r^2} dr^2
\end{equation}
where
\begin{eqnarray}
\label{def1}
{\cal F}&=& 1 - \sum_{i=1}^{N/2}
\frac{a_i^2 \mu_i^2}{(r^2 + a_i^2)}
                                        \\   \nonumber
\Pi&=& \prod_{i=1}^{N/2} (r^2 + a_i^2).
\end{eqnarray}
In Eq.(\ref{spacetime1}) we assumed that $N$ is even.  
The odd $N$ case will be discussed later.
The $\mu_i$ are not
all independent but obeys
\begin{equation}
\label{mu-rel}
\mu_1^2 + \mu_2^2 + \cdots + \mu_{N/2}^2 = 1.
\end{equation}
The mass $M$ and angular momenta $J_i$ of the black hole (\ref{spacetime1})
are 
\begin{equation}
\label{mass-angular}
M = \frac{(N-1) \Omega_{N-1}}{16 \pi G} \mu
\hspace{2.0cm}
J_i = \frac{2}{N-1} M a_i
\hspace{0.3cm}
(i = 1, 2, \cdots, \frac{N}{2})
\end{equation}
where $\Omega_{N-1} = 2 \pi^{N/2} / \Gamma[N/2]$ is the area of a unit 
$(N-1)$-sphere and $G$ is a $(N+1)$-dimensional Newton constant, which will
be assumed to be unity from now on.

Now, we would like to calculate the horizon area $A$ of the spacetime 
(\ref{spacetime1}), which is given by
\begin{equation}
\label{area1}
A = \int_0^{2 \pi} d\phi_1 \cdots d\phi_{\frac{N}{2}}
    \int_0^1 d \mu_1
    \int_0^{\sqrt{1 - \mu_1^2}} d \mu_2
    \cdots
    \int_0^{\sqrt{1 - \mu_1^2 - \cdots - \mu_{\frac{N}{2} - 2}^2}}
    d \mu_{\frac{N}{2} - 1}
    \sqrt{\mbox{det} M}
\end{equation}
where $\mbox{det} M = \mbox{det} (g_{\mu_i, \mu_j}) \mbox{det} (g_{\phi_i, 
\phi_j}) |_{r = r_H}$. The horizon radius $r_H$ is defined by solving 
\begin{equation}
\label{hori-rad}
\prod_{i=1}^{N/2} (r_H^2 + a_i^2) = \mu r_H^2.
\end{equation}
The factorization of $\mbox{det} M$ comes from $g_{\mu_i, \phi_j}=0$. It
is not difficult to show $\mbox{det} M = \mu^2 r_H^2 \mu_1^2 \mu_2^2 \cdots
\mu_{N/2 - 1}^2$, which makes $A$ in the following simple form
\begin{equation}
\label{area2}
A = \Omega_{N-1} \mu r_H.
\end{equation}

Now, we regard $M$ and $J_i (i=1, 2, \cdots, N/2)$ are independent variables.
Then one can write
\begin{equation}
\label{elementary}
dA = \frac{\partial A}{\partial M} d M + \sum_{i=1}^{N/2}
     \frac {\partial A}{\partial J_i} d J_i.
\end{equation}
Firstly, let us compute $\partial A / \partial M$, which is
\begin{equation}
\label{papm-1}
\frac{\partial A}{\partial M} = \Omega_{N-1}
\left[ \frac{\mu}{M} r_H + \mu \frac{\partial r_H}{\partial M} \right].
\end{equation}
To compute $\partial r_H / \partial M$ we use Eq.(\ref{hori-rad}). 
Differentiating Eq.(\ref{hori-rad}) with respect to $M$, one can show easily
\begin{equation}
\label{prhpm-1}
\frac{\partial r_H}{\partial M} = \frac{1}{{\cal B}}
\left[\frac{r_H}{2 M} + \frac{r_H}{M}
      \sum_{i=1}^{N/2}
      \frac{a_i^2}{r_H^2 + a_i^2}  \right]
\end{equation}
where
\begin{equation}
\label{deno-1}
{\cal B} = \sum_{i=1}^{N/2} \frac{r_H^2}{r_H^2 + a_i^2} - 1.
\end{equation}
It is important to note that ${\cal B}$ can be expressed as 
\begin{equation}
\label{calb-1}
{\cal B} = r_H \kappa
\end{equation}
where $\kappa$ is a surface gravity defined
\begin{equation}
\label{surface-1}
\kappa = \frac{\partial_r \Pi - 2 \mu r}{2 \mu r^2} \Bigg|_{r = r_H}.
\end{equation}
Since the surface gravity is proportional to the Hawking temperature, we 
can assume ${\cal B} > 0$.

Inserting Eq.(\ref{prhpm-1}) into (\ref{papm-1}) and using 
Eq.(\ref{mass-angular}) yields
\begin{equation}
\label{papm-2}
\frac{\partial A}{\partial M} = \frac{8 \pi r_H}{{\cal B}}.
\end{equation}
 
Next, let us compute $\partial A / \partial J_j$, which is 
\begin{equation}
\label{papj-1}
\frac{\partial A}{\partial J_j} = \Omega_{N-1} \mu 
\frac{\partial r_H}{\partial J_j}.
\end{equation}
Differentiating Eq.(\ref{hori-rad}) with respect to $J_j$, one can show easily
\begin{equation}
\label{prhpj-1}
\frac{\partial r_H}{\partial J_j} = - 
\frac{(N-1) r_H}{2 M {\cal B}} \Omega_j
\end{equation}
where 
\begin{equation}
\label{freq-1}
\Omega_j = \frac{a_j}{r_H^2 + a_j^2}
\end{equation}
is a frequency of the black hole arising due to the $j$-th angular momentum
$J_j$. Inserting Eq.(\ref{prhpj-1}) into (\ref{papj-1}) yields
\begin{equation}
\label{papj-2}
\frac{\partial A}{\partial J_j} = - \frac{8 \pi r_H}{{\cal B}}
\Omega_j.
\end{equation}
Thus inserting (\ref{papm-2}) and (\ref{papj-2}) into (\ref{elementary}) simply
yields
\begin{equation}
\label{thermo-1}
dA = \frac{8 \pi r_H}{{\cal B}}
\left[ dM - \sum_{i=1}^{N/2} \Omega_i dJ_i \right].
\end{equation}

Bekenstein has shown in Ref.\cite{beken73} that for scalar, electromagnetic, 
and gravitational waves $dJ_i / d M$ is expressed in terms of the stress-energy
tensor $T_{\mu \nu}$ as following
\begin{equation}
\label{se-tensor-1}
\frac{d J_i}{d M} = - 
\frac{T^r_{\phi_i}}{T^r_t}
\end{equation}
where $\phi_i$ are the azimuthal angles associated with $J_i$. Since the 
incident waves should have the factorization factors 
$e^{i m_i \phi_i} e^{-i \omega t}$, one can show easily that 
Eq.(\ref{se-tensor-1}) reduces to 
\begin{equation}
\label{se-tensor-2}
\frac{d J_i}{d M} = \frac{m_i}{\omega}
\end{equation}
where $m_i$ and $\omega$ are the azimuthal quantum numbers corresponding to 
$\phi_i$ and energy of the incident waves respectively. Inserting 
(\ref{se-tensor-2}) into (\ref{thermo-1}), one can show easily 
\begin{equation}
\label{thermo-2}
dA = \frac{8 \pi r_H}{{\cal B}} d M
\left[1 - \frac{1}{\omega} \sum_{i=1}^{N/2} m_i \Omega_i \right].
\end{equation}

Since $A/4$ is a black hole entropy, Eq.(\ref{thermo-2}) gives a 
condition 
\begin{equation}
\label{entro-1}
d M \left[ 1 - \frac{1}{\omega} \sum_{i=1}^{N/2} m_i \Omega_i \right] > 0.
\end{equation}
Since the existence of the superradiance modes implies $dM < 0$, it is 
easy to show that the condition for the existence of the superradiance is 
\begin{equation}
\label{superrad-1}
0 < \omega < \sum_{i=1}^{N/2} m_i \Omega_i.
\end{equation}
In Ref.\cite{frol03-1} the condition for the superradiance for the incident
scalar wave was shown to be $0 < \omega < m \Omega_a + k \Omega_b$, which is 
manifestly special case of Eq.(\ref{superrad-1}) in $N = 4$. Furthermore, 
our conclusion (\ref{superrad-1}) holds not only for the scalar wave but also
for the electromagnetic and gravitational waves.  

For a completeness we consider the odd $N$ case. In this case the metric
for the rotating black hole is changed into
\begin{eqnarray}
\label{spacetime2}
ds^2&=&-dt^2 + r^2 d\alpha^2 + \sum_{i=1}^{(N-1)/2} (r^2 + a_i^2)
(d \mu_i^2 + \mu_i^2 d \phi_i^2) 
                                 \\   \nonumber
& &\hspace{1.0cm} + 
\frac{\mu r}{\Pi {\cal F}}
\left[dt + \sum_{i=1}^{(N-1)/2} a_i \mu_i^2 d \phi_i \right]^2
+ \frac{\Pi {\cal F}}{\Pi - \mu r} d r^2
\end{eqnarray}
where
\begin{eqnarray}
\label{def11}
\Pi&=& \prod_{i=1}^{(N-1)/2} (r^2 + a_i^2)
                                           \\   \nonumber
{\cal F}&=& 1 - \sum_{i=1}^{(N-1)/2}
\frac{a_i^2 \mu_i^2}{r^2 + a_i^2}
\end{eqnarray}
and 
\begin{equation}
\label{mu-rel-11}
\sum_{i=1}^{(N-1)/2} \mu_i^2 + \alpha^2 = 1.
\end{equation}
The mass $M$ and the angular momenta $J_i$ of the black hole (\ref{spacetime2})
are same with Eq.(\ref{mass-angular}). The horizon area is slightly different
from Eq.(\ref{area2}):
\begin{eqnarray}
\label{area-11}
A&=& \int_0^{2\pi} d\phi_1 \cdots d\phi_{\frac{N-1}{2}}
\int_0^1 d\mu_1
\int_0^{\sqrt{1 - \mu_1^2}} d\mu_2  \cdots
\int_0^{\sqrt{1 - \mu_1^2 - \cdots - \mu_{\frac{N-3}{2}}^2}}
d \mu_{\frac{N-1}{2}}
                                        \\   \nonumber
& & \hspace{4.0cm} \times
\sqrt{\mbox{det} (g_{\mu_i \mu_j}) \bigg|_{r = r_H}
      \mbox{det} (g_{\phi_i \phi_j}) \bigg|_{r = r_H}
     }
                                         \\   \nonumber
&=& \frac{\Omega_{N-1}}{2} \mu r_H.
\end{eqnarray}
Then it is straightforward to show 
\begin{equation}
\label{mibun-1}
\frac{\partial A}{\partial M} = \frac{8 \pi r_H}{{\cal B}}
\hspace{1.0cm}
\frac{\partial A}{\partial J_i} = - \frac{8 \pi r_H}{{\cal B}} \Omega_i
\end{equation}
where $\Omega_i = a_i / (r_H^2 + a_i^2)$ are the rotational frequency
of the black hole and 
\begin{equation}
\label{deno-11}
{\cal B} = \frac{2 r_H}{\mu} \sum_{i=1}^{(N-1)/2}
\prod_{j \neq i} (r_H^2 + a_j^2) - 1.
\end{equation}
One can show easily ${\cal B} = 2 r_H \kappa$ where $\kappa$ is a surface
gravity defined
\begin{equation}
\label{surface-11}
\kappa = \frac{\partial_r \Pi - \mu}{2 \mu r} \Bigg|_{r = r_H}.
\end{equation}

Then it is easy to show 
\begin{equation}
\label{thermo-11}
dA = \frac{8\pi r_H}{{\cal B}}
\left[ d M - \sum_{i=1}^{(N-1)/2} \Omega_i d J_i \right].
\end{equation}
Following the same procedure we can conclude that the condition for the 
existence of the superradiance modes for the incident scalar, 
electromagnetic and gravitational waves reduces to 
\begin{equation}
\label{superrad-11}
0 < \omega < \sum_{i=1}^{(N-1)/2} m_i \Omega_i
\end{equation}
where $m_i$ are the azimuthal quantum numbers.

In this paper we derived the condition for the existence of the superradiance
modes for the incident scalar, electromagnetic and gravitational waves when the
spacetime background is a $(N+1)$-dimensional rotating black holes with 
multiple angular momentum parameters. Our final condition reduces to a 
simple form $0 < \omega < \sum_i m_i \Omega_i$, where $m_i$ is an azimuthal
quantum numbers of the incident waves associated with the $i$-th angle $\phi_i$
and $\Omega_i$ is a rotating frequency corresponding to the $i$-th 
angular momentum $J_i$. In $4d$ Kerr black hole it is well-known that there
is no superradiance mode for the incident fermionic wave\cite{unruh73,chand83}.
It is of interest to check whether this property is maintained in the 
higher-dimensional rotating black hole background or not. Another interesting
point arising due to the existence of the superradiance modes in the
rotating brane-world black holes is that the standard claim 
`{\it the black holes radiate mainly on the brane}' is not obvious in this 
background. Thus, it is necessary to check which one is dominant
between the bulk-emission and the brane-emission by adopting an appropriate
numerical method. We hope to report this issue elsewhere.

\vspace{1cm}

{\bf Acknowledgement}:  
This work was supported by the Korea Research
Foundation under Grant (KRF-2003-015-C00109).


\begin{thebibliography}{99}
\bibitem{ark98-1} N. Arkani-Hamed, S. Dimopoulos and G. Dvali,
{\it The Hierarchy Problem and New Dimensions at a Millimeter}, 
Phys. Lett. {\bf B429} (1998) 263 [hep-ph/9803315].
\bibitem{anto98} L. Antoniadis, N. Arkani-Hamed, S. Dimopoulos and G. Dvali,
{\it New Dimensions at a Millimeter to a Fermi and Superstrings at a 
TeV}, Phys. Lett. {\bf B436} (1998) 257 [hep-ph/9804398]. 
\bibitem{rs99-1} L. Randall and R. Sundrum, {\it A Large Mass Hierarchy from a 
Small Extra Dimension}, 
Phys. Rev. Lett. {\bf 83} (1999) 3370 [hep-ph/9905221].
\bibitem{gidd02-1} S. B. Giddings and T. Thomas, {\it High energy colliders 
as black hole factories: The end of short distance physics}, Phys. Rev. 
{\bf D65} (2002) 056010 [hep-ph/0106219].
\bibitem{dimo01-1} S. Dimopoulos and G. Landsberg, {\it Black Holes at the 
Large Hadron Collider}, Phys. Rev. Lett. {\bf 87} (2001) 161602 
[hep-ph/0106295].
\bibitem{eard02-1} D. M. Eardley and S. B. Giddings, {\it Classical black hole 
production in high-energy collisions}, Phys. Rev. {\bf D66} (2002)
044011 [gr-qc/0201034].
\bibitem{stoj04} D. Stojkovic, {\it Distinguishing between the small ADD and
RS black holes in accelerators}, Phys. Rev. Lett. {\bf 94} (2005)
011603 [hep-ph/0409124].
\bibitem{jung05-2} E. Jung and D. K. Park, {\it Absorption and Emission
Spectra of an higher-dimensional Reissner-Nordstr\"{o}m black hole}
[hep-th/0502002].
\bibitem{jung05-1} E. Jung, S. H. Kim and D. K. Park, {\it Ratio of 
absorption cross section for Dirac fermion to that for scalar in the
higher-dimensional black hole background} [hep-th/0503027].
\bibitem{unruh76} W. G. Unruh, {\it Absorption cross section of small 
black holes},
Phys. Rev. {\bf D14} (1976) 3251.
\bibitem{argy98} P. Argyres, S. Dimopoulos and J. March-Russell, 
{\it Black Holes and Sub-millimeter Dimensions}, Phys. Lett. 
{\bf B441} (1998) 96 [hep-th/9808138].
\bibitem{banks99} T. Banks and W. Fischler, {\it A Model for High Energy 
Scattering
in Quantum Gravity} [hep-th/9906038].
\bibitem{emp00} R. Emparan, G. T. Horowitz and R. C. Myers, {\it Black Holes
radiate mainly on the Brane}, Phys. Rev. Lett. {\bf 85} (2000) 499
[hep-th/0003118].
\bibitem{zeldo71-1} Y. B. Zel'dovich, {\it Generation of waves by a 
rotating body}, JETP Lett. {\bf 14} (1971) 180.
\bibitem{press72} W. H. Press and S. A. Teukolsky, {\it Floating Orbits, 
Superradiant Scattering and the Black-hole Bomb}, Nature {\bf 238} (1972) 211.
\bibitem{star73-1} A. A. Starobinskii, {\it Amplification of waves during 
reflection from a rotating black hole}, Sov. Phys. JETP {\bf 37} (1973)
28.
\bibitem{star74-1} A. A. Starobinskii and S. M. Churilov, {\it Amplification
of electromagnetic and gravitational waves scattered by a rotating black
hole}, Sov. Phys. JETP {\bf 38} (1974) 1.
\bibitem{frol02-1} V. Frolov and D. Stojkovi\'{c}, {\it Black hole radiation
in the brane world and the recoil effect}, Phys. Rev. {\bf D66} (2002)
084002 [hep-th/0206046].
\bibitem{frol02-2} V. Frolov and D. Stojkovi\'{c}, {\it Black Hole as a Point
Radiator and Recoil Effect on the Brane World}, Phys. Rev. Lett. 
{\bf 89} (2002) 151302 [hep-th/0208102].
\bibitem{frol03-1} V. Frolov and D. Stojkovi\'{c}, {\it Quantum radiation from 
a $5$-dimensional black hole}, Phys. Rev. {\bf D67} (2003) 084004
[gr-qc/0211055].
\bibitem{ida05} D. Ida, K. Oda and S. C. Park, {\it Anisotropic scalar field
emission from TeV scale black hole} [hep-ph/0501210].
\bibitem{harris05-1} C. M. Harris and P. Kanti, {\it Hawking Radiation from a 
$(4+n)$-Dimensional Rotating Black Hole} [hep-th/0503010].
\bibitem{jung05-3} E. Jung, S. H. Kim and D. K. Park, {\it Condition for 
Superradiance in Higher-dimensional Rotating Black Holes}
[hep-th/0503163].
\bibitem{beken73} J. D. Bekenstein, {\it Extraction of Energy and Charge from 
a Black Hole}, Phys. Rev. {\bf D7} (1973) 949.
\bibitem{myers86} R. C. Myers and M. J. Perry, {\it Black Holes in Higher
Dimensional Space-Times}, Ann. Phys. {\bf 172} (1986) 304.
\bibitem{unruh73}W. Unruh, {\it Separability of the Neutrino Equations in a 
Kerr Background}, Phys. Rev. Lett. {\bf 31} (1973) 1265.
\bibitem{chand83} S. Chandrasekhar, {\it The Mathematical Theory of Black Hole}
(Oxford University Press, New York, 1983).
\end{thebibliography}
\end{document}